\documentclass{an}
\usepackage{graphicx}
\usepackage{times}
\Volume{000}             
\Year{0000}              
\Month{00}               
\Pagespan{000}{000}      

\begin{document}
\lhead[\thepage]{B. Stelzer et al.: Coordinated Multiwavelength Observations of V410\,Tau}
\rhead[Astron. Nachr./AN~{\bf XXX} (200X) X]{\thepage}
\headnote{Astron. Nachr./AN {\bf 32X} (200X) X, XXX--XXX}

\title{Coordinated Multiwavelength Observations of V410\,Tau}

\author{B. Stelzer$^1$, M. Fern\'andez$^2$, V. Costa$^3$, K. Grankin$^4$, A. Henden$^5$, J. F. Gameiro$^3$, E. Guenther$^6$, R. Jayawardhana$^7$, S. Mohanty$^7$, V. Burwitz$^1$, P. Predehl$^1$, R. H. Durisen$^8$}
\institute{$^1$ Max-Planck-Institut f\"ur extraterrestrische Physik, Postfach 1312, D-85741 Garching, Germany, \newline $^2$ Instituto de Astrof\'\i sica de Andaluc\'\i a, Camino Bajo de Hu\'etor 24, E-18008 Granada, Spain, \newline $^3$ Centre for Astrophysics, University of Porto, Rua das Estrelas, 4150 Porto, Portugal, \newline $^4$ Ulug Beg Astronomical Institute, Astronomicheskaya 33, 700052 Tashkent, Uzbekistan, \newline $^5$ USRA/USNO Flagstaff Station, P. O. Box 1149, Flagstaff, AZ 86002-1149, USA, \newline $^6$ Th\"uringer Landessternwarte, Karl-Schwarzschild-Observatorium, Sternwarte 5, D-07778 Tautenburg, Germany, \newline $^7$ Astronomy Department, University of California, Berkeley, CA 94720, USA, \newline $^8$ Indiana University, Swain West 319, Bloomington, IN 47405, USA}
\date{Received {\it date will be inserted by the editor}; 
accepted {\it date will be inserted by the editor}} 

\abstract{
In November 2001 we undertook a coordinated observing campaign to study the connection between X-ray and optical variability in the weak-line T Tauri star V410\,Tau. The observing plan included three $\sim 15$\,ksec observations with {\em Chandra} using the Advanced CCD Imaging Spectrometer for Spectroscopy (ACIS-S) scheduled for different phases of the known 1.87\,d starspot cycle. Photometric and spectroscopic monitoring of V410\,Tau involving telescopes on three different continents was scheduled simultaneously with the {\em Chandra} exposures.
}
\correspondence{stelzer@mpe.mpg.de}

\maketitle

\section{Introduction}\label{sect:intro}

V410\,Tau is a well-known T Tauri star (TTS) of the weak-line class. 
Due to its optical brightness ($V \sim 11$\,mag) V410\,Tau has 
frequently been included in photometric monitoring programs throughout the last
three decades. These observations have revealed periodic variability 
on a $1.87$\,d cycle, which has remained stable over many years
(e.g. Vrba et al. 1988, Herbst 1989, Petrov et al. 1994 hereafter P94). 
The periodic behavior of V410\,Tau 
was attributed to surface spots similar to those 
observed on BY\,Dra and RS\,CVn systems. 
The amplitude of the asymmetric lightcurve of V410\,Tau 
is larger than that of any other weak-line TTS and varies with a time-scale of
years between $0.2-0.6$\,mag in the $V$ band.

Optical photometric and spectroscopic monitoring has occasionally resulted 
in the detection of flares (Vrba et al. 1988, Guenther \& Ball 1999). However,
in view of the large existing data base for V410\,Tau, flares are surprisingly
rare on this star. 
This may indicate that the magnetic field of V410\,Tau maintains a  
comparatively stable configuration, an idea supported also by the long-term 
stability of the photometric spot cycle.
On the other hand V410\,Tau is a strong and variable X-ray source
(Costa et al. 2000) suggesting that magnetic fields are dissipated in the
corona. Despite its obvious variability in the X-ray regime, 
no direct signs of X-ray flares have been observed on V410\,Tau so far. 

In order to study the connection between the optical and X-ray variability 
of V410\,Tau we have undertaken a coordinated observing campaign involving 
photometric monitoring, optical low- and high-resolution spectroscopy, 
and {\em Chandra} observations 
at different phases throughout the (photometric) spot cycle.
Here we report on the first results from the photometric and X-ray monitoring.

\section{Observations}\label{sect:observations}

In Nov 2001 we obtained simultaneous observations of V410\,Tau 
in the optical 
and X-ray regime. 

Three {\em Chandra} ACIS-S exposures 
of 15\,ksec each were scheduled for different phases of the known photometric
spot cycle. 
If the X-ray emission comes from spotted regions,  
the maximum of the X-ray emission should occur at the minimum of the optical
lightcurve, i.e. at phases when the spot is on the visible hemisphere, and 
vice versa. 
Two of the {\em Chandra} observations were performed in Nov 2001.
Due to high solar activity, the third observation had to be rescheduled, and
was carried out in March 2002 without optical coverage.

Optical photometric observations accompanying the {\em Chandra} data from 
Nov 2001 were carried out at three sites, the USNO in Flagstaff (Arizona), 
the Mt. Maidanak Observatory in Uzbekistan, and the Sierra Nevada Observatory
in Spain. 
In addition we performed spectroscopic observations at Sierra Nevada, 
Lick, and Calar Alto observatories. 
The spectroscopy will be discussed in a later publication.

\section{Phase-Folded Optical and X-ray Lightcurves}\label{sect:lcs}

\begin{figure*}
\begin{center}
\includegraphics[width=18cm, angle=0]{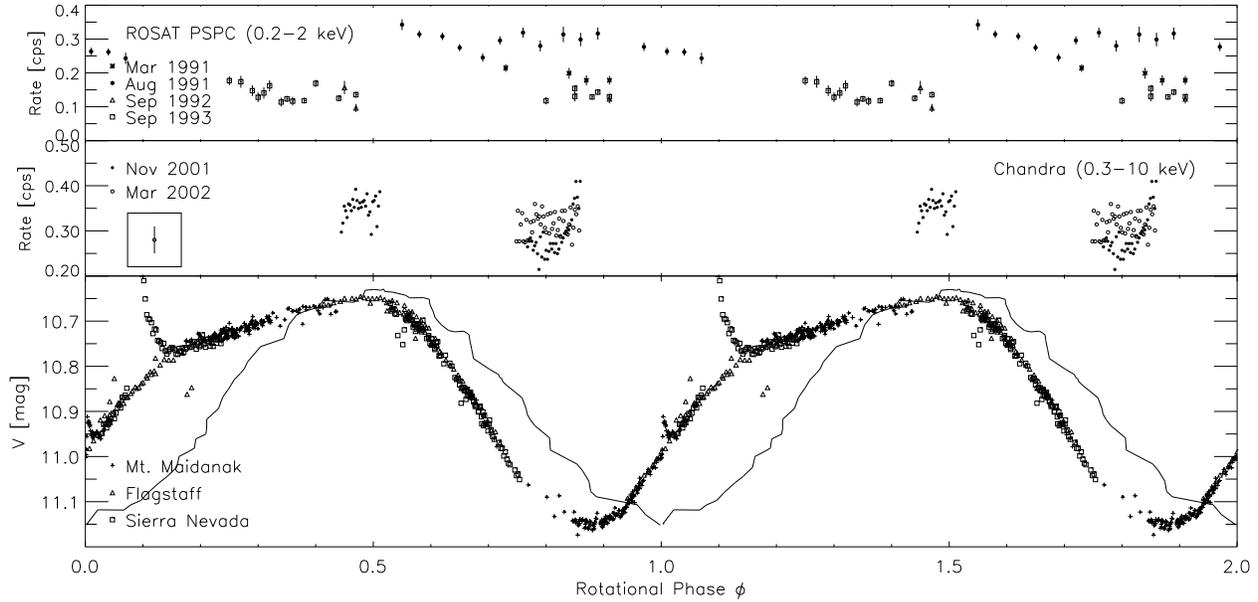}
\caption{Phase folded X-ray and optical lightcurve of V410\,Tau. {\em Lower panel} - The optical data are obtained in the course of our campaign in Nov\,2001. For comparison the $V$ band lightcurve observed in 1992 from Mt. Maidanak (Grankin 1999) is also shown (solid curve). {\em Middle panel} - Two of the {\em Chandra} observations were obtained quasi-simultaneous with the optical photometry. The third one was carried out in March 2002. Binsize is 400\,s. {\em Upper panel} - All available pointed {\em ROSAT} PSPC data for V410\,Tau were re-analysed and added to this figure to enhance the phase coverage of the X-ray data. Binsize is 400\,s.}
\label{fig:phase_lcs}
\end{center}
\end{figure*}

To examine the variability, we generated a lightcurve for
each of the three {\em Chandra} observations, and folded these lightcurves 
with the rotation period. 
In Fig.~\ref{fig:phase_lcs} we display the quasi-simultaneous {\em Chandra} 
X-ray and the combined $V$ band lightcurves of three optical observatories. 

All data shown in Fig.~\ref{fig:phase_lcs} were folded using the ephemeris 
given by P94.
A clear shift in the location of the optical minimum with respect to phase 
$\phi = 0.0$ is obvious. 
Under the assumption that the distribution of the spots is constant in time,
this migration of the minimum points at the need for a new 
determination of the rotation period (see Sect.~\ref{sect:ephemeris}). 
Changes in the spot pattern as compared to earlier measurements
(the lightcurve from 1992 is displayed for comparison 
in Fig.~\ref{fig:phase_lcs}; Grankin 1999) can be inferred
from the fact that the shape of the lightcurve changed.

The {\em Chandra} lightcurves of V410\,Tau are displayed in the 
middle panel. The systematic rise of the X-ray 
count rate near optical minimum in the observation 
from Nov 2001 is not repeated in the March 2002 observation, 
and it is therefore more likely that it is due to erratic variability, 
such as a flare, than due to a rotational effect.
No obvious correlation with the optical lightcurve is 
observed suggesting that on V410\,Tau rotational cycles are not visible in 
the X-ray regime. 

To enhance the phase coverage of the X-ray data we have computed and 
phase-folded lightcurves for all pointed 
{\em ROSAT} PSPC observations of V410\,Tau. These lightcurves are  
shown in the upper panel of Fig.~\ref{fig:phase_lcs}. 
Details about these observations can be found in Costa et al. (2000),
especially in their Table~4. Note that nearly ten years have passed
between the {\em ROSAT} and the {\em Chandra} observations. 
From the plot in Fig.~\ref{fig:phase_lcs} we can not see a correlation
between the {\em ROSAT} count rate and the optical spot cycle. 
This may be understood in analogy with the Sun, where the X-ray emission
does not coincide spatially with spotted regions. V410\,Tau
shows modest X-ray variability except in August 1991, where the source seems
to have undergone a phase of high activity, both in terms of count rate
and variability.

\section{A New Period Determination}\label{sect:ephemeris}

For the new determination of the rotation period of V410\,Tau, 
as a first step, the minimum of the $V$ band lightcurve from Nov 2001
was computed by approximating the data between $\phi = 0.73-0.99$ by a second
degree polynomial. The resulting phase of minimum ($\phi = 0.865 \pm 0.025$) 
converts to a new reference point $T_{\rm min}$ 
(see Table~\ref{tab:ephemeris}). 
Making use of the number of rotational cycles elapsed between the old minimum 
(observed by P94)
and the minimum observed in Nov 2001 we derived the updated period. 
The uncertainty in the new value for the period 
stems from the uncertainty in the determination of the minimum.
The two values for the period differ by more than 3\,$\sigma$, indicating
that either the uncertainty was underestimated by P94, or the spots have
migrated in longitude.
\begin{table}
\caption{Ephemeris of V410\,Tau: value by P94 and
new update.}
\label{tab:ephemeris}
\begin{center}
\begin{tabular}{llr} \hline
Reference & \multicolumn{1}{c}{$T_{\rm min}\,{\rm [JD]}$} & \multicolumn{1}{c}{$P_{\rm rot}\,{\rm [d]}$} \\
\hline
P94   & $2446659.4389$   & $1.872095 \pm 0.000022$ \\
here  & $2452234.285971$ & $1.872010 \pm 0.000016$ \\
\hline
\end{tabular}
\end{center}
\end{table}

\begin{figure}
\begin{center}
\resizebox{9cm}{!}{\includegraphics{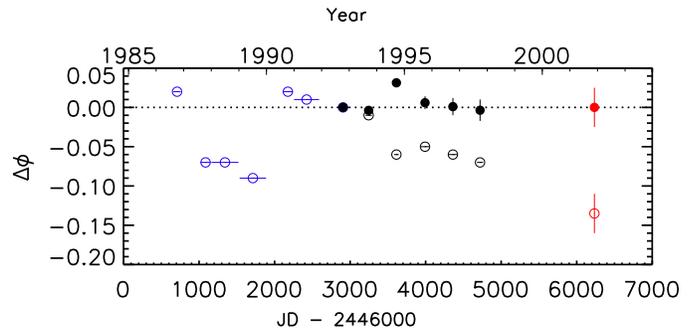}}
\caption{Phase shift $\Delta \phi$ of the minimum brightness of V410\,Tau with respect to the ephemeris given by P94 (open circles), and after correction with the new value for the rotation period (filled circles).}
\label{fig:migration}
\end{center}
\end{figure}

V410\,Tau has constantly but irregularly been monitored throughout the
last three decades. Among others extensive photometric observations 
in the time between 1992 (the previous determination of the ephemeris)
and our campaign in 2001 were carried out at Mt. Maidanak Observatory.
Grankin (1999) reports a gradual migration of the phase of minimum light
in their analysis of the $V$ band lightcurve of V410\,Tau using data 
from 1986 to 1997. This trend is continued by our observations 
(see Fig.~\ref{fig:migration}).
We folded the $V$ band lightcurves of V410\,Tau obtained in the time interval 
spanned by P94's determination of the minimum and our observations 
on a yearly basis using our new values for $T_{\rm min}$ and $P_{\rm rot}$. 
Fig.~\ref{fig:migration} 
shows the phase shifts of the minimum for seasonal averages of 
published $V$ band lightcurves with respect to P94 (open circles). 
and our new ephemeris (filled circles). 
Most of the averages are now compatible with $\Delta \phi = 0$
within their errors.

\section{A Large Optical Flare}\label{sect:flare}

On 24 Nov, 2001 V410\,Tau exhibited an optical flare,
which was observed from Sierra Nevada observatory
using Stroemgren $uvby$ filters (see Fig.~\ref{fig:phase_lcs}).
A similar though less pronounced brightness increase is seen in the
data from Mt. Maidanak near $\phi = 0.0$ in Fig.~\ref{fig:phase_lcs}.
Here, we discuss only the large flare observed from Sierra Nevada.

In the $v$ band most of the output is likely due to emission lines. 
The $b$ band includes no prominent lines, such that the observed 
variability probably represents variations in the continuum.
In both $v$ and $b$ bands the band-width (full width half maximum) 
is the same. Therefore, we can directly compare both lightcurves:
Although the amplitude in the $b$ band is smaller than in the $v$ band
(see Table~\ref{tab:flare}), the increase in temperature related to the flare
also produces substantial continuum emission. We computed the decay times 
in the $uvby$ bands, and found that the decay time prolongs 
towards shorter wavelengths along with an increase of the amplitude.
Table~\ref{tab:flare} 
gives the time required during the flare decay to fall back to 
50\,\% and 10\,\% of the peak for different bands.
Within the time-resolution of $\sim$\,6\,min no lags between the 
different bands are observed, i.e. the maximum occurred at $\sim$ UT 21:14
in all bands.

\begin{table}
\begin{center}
\caption{Characteristics of the flare on Nov 2001: amplitudes $\Delta$\,mag and decay times to 50\,\% and 10\,\% of the peak.}
\label{tab:flare}
\begin{tabular}{crrr} \hline 
Band & $\Delta$\,mag & \multicolumn{1}{c}{$t_{\rm dec}$} & \multicolumn{1}{c}{$t_{\rm dec}$} \\ 
     &               & \multicolumn{1}{c}{$50\,$\%}       & \multicolumn{1}{c}{$10$\,\%}       \\
\hline
   $u$ &       3.04    &  1h 12m 51s & 3h 27m 57s \\   
   $v$ &       2.04    &  0h 45m 19s & 2h 41m 22s \\   
   $b$ &       1.32    &  0h 27m 57s & 1h 43m 46s \\   
   $y$ &       0.95    &  0h 28m 48s & 1h 41m 39s \\   
\hline
\end{tabular}
\end{center}
\end{table}

\section{Summary}\label{sect:summary}

The simultaneous X-ray and optical data of this campaign does not support a direct
connection between X-ray emitting regions and starspots on V410\,Tau. In contrast to most
previous observations of this star, our continuous monitoring has resulted
in the detection of at least one large flare. We tentatively re-determined the rotation
period based on the migration of the minimum in the $V$ band lightcurve. Future analysis
will show whether this migration could instead be due to a shift of the spot pattern on
the surface of V410\,Tau.

\end{document}